\documentclass[]{article}

\usepackage[titletoc,toc,title]{appendix}
\usepackage{graphicx}
\usepackage{dcolumn}
\usepackage{bm}
\usepackage{subfig}
\usepackage{caption}
\usepackage{epsfig}
\usepackage{epstopdf}
\usepackage[numbers,sort&compress]{natbib}
\usepackage{graphics}
\usepackage{latexsym}
\usepackage{amsmath}
\usepackage{calrsfs} 
\usepackage{rotating}
\usepackage{authblk}
\usepackage{xfrac}
\usepackage{color}

\newcommand{\eg}{{\it e.g.}}

\newcommand*\samethanks[1][\value{footnote}]{\footnotemark[#1]}

\graphicspath{{./pics/}}
\date{\today}

\title{Catenoid stability with a free contact line} 

\author{Amir Akbari\thanks{Department of Chemical Engineering, McGill University, Montreal, Quebec H3A 0C5} \and Reghan J. Hill\samethanks \and Theo G.M. van de Ven\thanks{Department of Chemistry, McGill University, Montreal, Quebec H3A 2A7}}

\begin{document}

\maketitle

\begin{abstract}
Contact-drop dispensing is central to many small-scale applications, such as direct-scanning probe lithography and micromachined fountain-pen techniques. Accurate and controllable dispensing required for nanometer-resolved surface patterning hinges on the stability and breakup of liquid bridges. Here, we analytically study the stability of catenoids pinned at one contact line with the other free to move on a substrate subject to axisymmetric and non-axisymmetric perturbations. We apply a variational formulation to derive the corresponding stability criteria. The maximal stability region and stability region are represented in the favourable and canonical phase diagrams, providing a complete description of catenoid equilibrium and stability. All catenoids are stable with respect to non-axisymmetric perturbations. For a fixed contact angle, there exists a critical volume below which catenoids are unstable to axisymmetric perturbations. Equilibrium solution multiplicity is discussed in detail, and we elucidate how geometrical symmetry is reflected in the maximal stability and stability regions. 
\end{abstract}



\section{Introduction} \label{sec:introduction}

The celebrated treatise of \citet{plateau1873statique} was a key study in the nineteen century, addressing liquid-bridge stability and breakup. Early investigations were motivated by applications such as liquid-jet breakup \citep{rayleigh1879capillary, tomotika1935instability}, crystal growth in microgravity \citep{velarde1988physicochemical}, oil recovery \citep{smith1985separation}, and paper wet strength \citep{tejado2010does}. Recently, interests have grown into areas such as elastocapillarity \citep{kwon2008equilibrium, taroni2012multiple, duprat2012wetting, akbari2014an}, contact-drop dispensing \citep{qian2009micron} with applications to scanning-probe lithography \citep{liu2000nanofabrication} and micromachined fountain-pen techniques \citep{deladi2004micromachined}. Molecular-resolution surface patterning provides new opportunities for advanced tissue engineering \citep{gadegaard2006applications}, DNA self-assembled nanoconstructs \citep{shen2009nta}, and highly sensitive protein chips \citep{choi2009fabrication}.

Static stability analysis of liquid bridges can be traced to the nineteenth century \citep{plateau1873statique, howe1887rotations}. Howe's variational formulation extended Plateau's primitive theory for cylindrical interfaces to unbounded axisymmetric capillary surfaces subject to a constant-volume. His criteria (sufficient conditions for the weak extrema of a functional) guarantee a surface to have the minimum energy among all the neighbouring surfaces of revolution. \citet{gillette1971stability} applied Howe's method to predict the stability limit of bounded axisymmetric liquid bridges with respect to axisymmetric perturbations. These criteria were later generalized for arbitrary interfaces with arbitrary perturbations \citep{myshkis1987low}.

Catenoids are doubly-connected surfaces of revolution with zero mean curvature. They are special cases of constant-mean-curvature axisymmetric surfaces, and are important to stability studies on weightless liquid bridges for two reasons: (1) Stability criteria can be obtained analytically for catenoids, which helps guide numerical algorithms for general liquid brides in the small pressure (mean curvature) limit, and (2) the curve corresponding to catenoidal interfaces in the volume-slenderness phase diagram defines a boundary between regions of positive and negative capillary pressure \citep{myshkis1987low}. This is important for mechanical systems with dynamics that are driven by capillary pressure (\eg, elasto-capillary systems \citep{mastrangelo1993mechanical}). Previous studies have considered catenoids bridging two circular discs of the same radius \citep{erle1970stability}, catenoids between two parallel plates with both contact lines free to move \citep{strube1992stability, langbein1992stability, zhou1997stability}, and catenoids between a plat and sphere \citep{orr1975pendular}. However, these results are not applicable to contact-drop dispensing applications where the liquid forms a bridge with a free contact line at one end.

 {\color{black}
Recent studies on contact-drop dispensing have shown that the deposited drop size can be adjusted by the needle retraction speed, needle-tip size, surface characteristics, and dispensing control parameters \cite{qian2009micron, qian2011motion}. However, these studies do not distinguish the effect of static parameters from dynamic ones. A static stability analysis of liquid bridges with a free contact line furnishes the maximum-height stability limit, which reasonably approximates the pinch-off height at small capillary numbers $\mbox{Ca}\ll 1$ (quasi-static limit) \citep{eggers1997nonlinear, dodds2009stretching}. Critical perturbations estimate the dispensed drop volume and show how the bridge dynamically evolves. Here, we only focus on the catenoid as an important special case since equilibrium solution multiplicity and stability criteria can be determined analytically.}

In this paper, we analytically study the static stability of catenoids pinned at one contact line with the other free to move on a flat substrate with respect to constant-volume perturbations. This furnishes a two-dimensional phase diagram in which the stability region is represented with respect to the catenoid volume and slenderness. The effect of the catenoid geometrical symmetry on the stability region boundaries is discussed. We also present a phase diagram with respect to canonical variables, which facilitates the representation of symmetry in the stability region, maximal stability region, and multiple equilibrium solutions subject to various constraints. \citet{myshkis1987low} described how free contact lines are generally treated in their variational method. However, the stability criteria were not presented for liquid bridges with free contact lines. Therefore, we first present an exposition of Myshkis's variational formulation \citep{myshkis1987low}, and then derive the stability criteria in section~\ref{sec:theory} for axisymmetric liquid bridges with a free contact line. Equilibrium solution multiplicity is discussed in section~\ref{sec:equilibrium}, and the maximal stability and stability regions are determined for cylinders and catenoids in sections~\ref{sec:stability:cylinder} and \ref{sec:stability:catenoid}, respectively. The results are summarized in sections~\ref{sec:conclusion}.

\section{Theory} \label{sec:theory}

\begin{figure} 
\centering
\includegraphics[width=\linewidth]{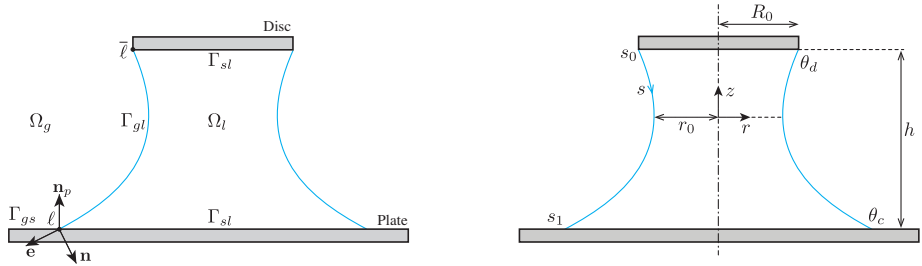}
 \caption{Catenoidal liquid bridge: schematic (left) and coordinate system with meridian curve parametrization (right).}
\label{fig:figure1}
 \end{figure}
 
We consider a liquid of volume $v$ bridging a circular disk with radius $R_{0}$ and a large plate. The disc and plate are separated by a distance $h$, as shown in Fig.~\ref{fig:figure1}. The region occupied by the liquid bridge is denoted $\Omega_{l}$, and that occupied by the surrounding fluid (of a different phase) is denoted $\Omega_{g}$. The bridge is pinned to the disc and free to slide horizontally on the plate. We restrict our analysis to catenoidal liquid bridges, which implies that the regions $g$ and $l$ have the same density and pressure, and the interface $\Gamma_{gl}$ is a surface of revolution. The formulation is presented as the limit of axisymmetric weightless liquid bridges with mean curvature approaching zero. The origin of the coordinate system is placed on the plane passing through the catenoid neck such that the $z$-axis is the symmetry axis. The meridian curve is parametrized with respect to its arclength $s$ such that $s=0$ at $z=0$. An equilibrium surface is specified by
\begin{equation}
	\left\{ \begin{array}{l}
         		r=r(s)\\
		z=z(s)\\
	\end{array} \right. 
	 \qquad s\in[s_{0},s_{1}],
    	 \label{eqn:eq1}
\end{equation}
extremizing the potential energy 
\begin{equation}
	U=\gamma_{sl}\Gamma_{sl}+\gamma_{gl}\Gamma_{gl}+\gamma_{sg}\Gamma_{sg},
    	 \label{eqn:eq2}
\end{equation}
where $\gamma_{ij}$ is the surface tension between the phases $i$ and $j$, and $\Gamma_{ij}$ is the interfacial surface area. Following \citet{myshkis1987low}, this leads to the Young-Laplace equation
\begin{equation}
	\left\{ \begin{array}{l}
         		r''=-z'(q-z'/r)\\
		z''=r'(q-z'/r)\\
	\end{array} \right. 
	 \qquad ('\equiv \mbox{d}/\mbox{d}s)
    	 \label{eqn:eq3}
\end{equation}
for axisymmetric equilibrium surfaces and
\begin{equation}
	\gamma_{gl}\cos\theta_{c}=\gamma_{sg}-\gamma_{sl}, \qquad \mbox{and} \qquad \cos\theta_{c}=\textbf{n}\cdot\textbf{n}_{p},
    	 \label{eqn:eq4}
\end{equation}
 where $q=-2k_{m}$ and $\theta_{c}$ is the contact angle. Here, $k_{m}$ is the mean curvature, which is zero for catenoids. Equation~(\ref{eqn:eq4}) shows that the contact angle is a thermodynamic property of the three-phase ($g$, $l$, and $s$) contact line, which is a constant for a specific substrate (plate) and the fluids occupying $\Omega_{g}$ and $\Omega_{l}$. Note that the dihedral angle $\theta_{d}$ can vary independently with the bridge volume to extremize the potential energy. Introducing the following lengths, which are scaled with the neck radius,
 \begin{equation}
	\hat{r}=r/r_{0}, \quad \hat{z}=z/r_{0}, \quad \hat{s}=s/r_{0},
    	 \label{eqn:eq5}
\end{equation}
the cylindrical volume $V=v/(\pi R_{0}^{2}h)$ and slenderness $\Lambda=h/R_{0}$ are two dimensionless parameters with which to present the phase diagram.

Following the method of \citet{myshkis1987low}, the interface stability is determined by the sign of the second variation. Using the Ritz method \citep{gelfand2000calculus}, the second variation is associated with the eigenvalues of the corresponding Strum-Liouville problem. Stability studies are generally concerned with determining stability regions in the phase diagram. Stability-region boundaries, identified by $\delta^{2} U=0$, correspond to critical states, separating stable equilibrium surfaces from unstable ones. Hence, we seek the conditions where $\lambda_{0}$ or $\lambda_{1}=0$, resulting in
\begin{equation}
	\left\{ \begin{array}{l}
         		\mathcal{L}\varphi_{0}+\mu=0\\
		\varphi_{0}(\hat{s}_{0})=0, \quad \quad \varphi'_{0}(\hat{s}_{1})+\hat{\chi} \varphi_{0}(\hat{s}_{1})=0\\
		\int_{\hat{s}_{0}}^{\hat{s}_{1}} \hat{r} \varphi_{0}\mbox{d}\hat{s}=0\\
	\end{array} \right. 
    	 \label{eqn:eq12}
\end{equation}
for axisymmetric perturbations and 
\begin{equation}
	\left\{ \begin{array}{l}
         		(\mathcal{L}-1/\hat{r}^{2})\varphi_{1}=0\\
		\varphi_{1}(\hat{s}_{0})=0, \quad \quad \varphi'_{1}(\hat{s}_{1})+\hat{\chi} \varphi_{1}(\hat{s}_{1})=0\\
	\end{array} \right. 
    	 \label{eqn:eq13}
\end{equation}
for non-axisymmetric perturbations, where 
\begin{equation}
	\chi=\frac{k_{1\ell}\cos \theta_{c}-k_{p\ell}}{\sin \theta_{c}} \quad \mbox{at } \ell,
	 \label{eqn:eq7}
\end{equation}
\begin{equation}
	\mathcal{L} \equiv \frac{\mbox{d}^{2}}{\mbox{d}\hat{s}^{2}}+\frac{\hat{r}'}{\hat{r}}\frac{\mbox{d}}{\mbox{d}\hat{s}}+\left[ \left( \hat{q}-\frac{\hat{z}'}{\hat{r}}\right)^{2}+\left(\frac{\hat{z}'}{\hat{r}}\right)^{2}\right]
    	 \label{eqn:eq14}
\end{equation}
with $\hat{q}=qr_{0}$, $\hat{\chi}=r_{0}\chi$, and $k_{1\ell}$, $k_{p\ell}$ the first principal curvatures of the interface and plate at the contact line $\ell$. The solutions of Eqs.~(\ref{eqn:eq12}) and (\ref{eqn:eq13}) can be written
\begin{equation}
	\varphi_{0}(\hat{s})=C_{1}w_{1}(\hat{s})+C_{2}w_{2}(\hat{s})+\mu w_{3}(\hat{s}),
    	 \label{eqn:eq15}
\end{equation}
\begin{equation}
	\varphi_{1}(\hat{s})=C_{4}w_{4}(\hat{s})+C_{5}w_{5}(\hat{s})
    	 \label{eqn:eq16}
\end{equation}
for axisymmetric and non-axisymmetric perturbations, respectively \citep{myshkis1987low}. These satisfy the following differential equations and their initial conditions

\begin{equation}
	\mathcal{L}w_{1}=0,\qquad w_{1}(0)=0,\quad w'_{1}(0)=1,
    	 \label{eqn:eq17}
\end{equation}
\begin{equation}
	\mathcal{L}w_{2}=0,\qquad w_{2}(0)=1,\quad w'_{2}(0)=0,
    	 \label{eqn:eq18}
\end{equation}
\begin{equation}
	\mathcal{L}w_{3}+1=0,\qquad w_{3}(0)=-1/4,\quad w'_{3}(0)=0,
    	 \label{eqn:eq19}
\end{equation}
\begin{equation}
	(\mathcal{L}-1/\hat{r}^{2})w_{4}=0,\qquad w_{4}(0)=0,\quad w'_{4}(0)=1,
    	 \label{eqn:eq20}
\end{equation}
\begin{equation}
	(\mathcal{L}-1/\hat{r}^{2})w_{5}=0,\qquad w_{5}(0)=1,\quad w'_{5}(0)=0,
    	 \label{eqn:eq21}
\end{equation}
where $w_{1}$, $w_{4}$ are odd and $w_{2}$, $w_{3}$, $w_{5}$ are even functions. Note that the initial conditions in Eq.~(\ref{eqn:eq19}) can be arbitrarily chosen because they do not affect the conditions describing the critical states of equilibrium surfaces (Eqs.~(\ref{eqn:eq22}) and (\ref{eqn:eq25})). The homogeneous solution of Eq.~(\ref{eqn:eq19}) is obtained from a linear combination of $w_{1}$ and $w_2$. From Eq.~(\ref{eqn:eq15}), the homogeneous part of $w_{3}$ makes no independent contribution to the general solution of $\varphi_{0}$. Therefore, the initial conditions for $w_{3}$ are chosen such that the general solution for $w_{3}$ comprises only the particular part. 

The critical state of an equilibrium surface is identified by the existence of a non-trivial solution for $\varphi_{0}$ or $\varphi_{1}$. These existence conditions can be obtained from Eqs.~(\ref{eqn:eq15}) and (\ref{eqn:eq16}) as
\begin{equation}
	\hat{\chi}^{0}=
	-\frac
	{\left| \begin{array} {c c c}
	w_{1}(\hat{s}_{0}) & w_{2}(\hat{s}_{0}) & w_{3}(\hat{s}_{0})\\
	w'_{1}(\hat{s}_{1}) & w'_{2}(\hat{s}_{1}) & w'_{3}(\hat{s}_{1})\\
	\int_{\hat{s}_{0}}^{\hat{s}_{1}} \hat{r} w_{1}\mbox{d}\hat{s} & \int_{\hat{s}_{0}}^{\hat{s}_{1}} \hat{r} w_{2}\mbox{d}\hat{s} & \int_{\hat{s}_{0}}^{\hat{s}_{1}} \hat{r} w_{3}\mbox{d}\hat{s}
	\end{array} \right|}
    	{\left| \begin{array} {c c c}
	w_{1}(\hat{s}_{0}) & w_{2}(\hat{s}_{0}) & w_{3}(\hat{s}_{0})\\
	w_{1}(\hat{s}_{1}) & w_{2}(\hat{s}_{1}) & w_{3}(\hat{s}_{1})\\
	\int_{\hat{s}_{0}}^{\hat{s}_{1}} \hat{r} w_{1}\mbox{d}\hat{s} & \int_{\hat{s}_{0}}^{\hat{s}_{1}} \hat{r} w_{2}\mbox{d}\hat{s} & \int_{\hat{s}_{0}}^{\hat{s}_{1}} \hat{r} w_{3}\mbox{d}\hat{s}
	\end{array} \right|},
	 \label{eqn:eq22}
\end{equation}

\begin{equation}
	\hat{\chi}^{1}=
	-\frac
	{\left| \begin{array} {c c }
	w_{4}(\hat{s}_{0}) & w_{5}(\hat{s}_{0})\\
	w'_{4}(\hat{s}_{1}) & w'_{5}(\hat{s}_{1})
	\end{array} \right|}
    	{\left| \begin{array} {c c }
	w_{4}(\hat{s}_{0}) & w_{5}(\hat{s}_{0})\\
	w_{4}(\hat{s}_{1}) & w_{5}(\hat{s}_{1})
	\end{array} \right|}.
	 \label{eqn:eq23}
\end{equation}
Here, $\hat{\chi}^{0}$ and $\hat{\chi}^{1}$ are the critical $\hat{\chi}$ corresponding to axisymmetric and non-axisymmetric perturbations, respectively. Note that $\hat{\chi}=$max$\{\hat{\chi}^{0},\hat{\chi}^{1}\}$ identifies a critical state. For a fixed $\Gamma_{gl}$, the minimum eigenvalue of the Sturm-Liouville problem is monotonically increasing with $\chi$. Hence, $\lambda_{i}>0$ for $\chi>\chi^{i}$ ($i=0,1$). It follows that an equilibrium surface is unstable with respect to axisymmetric (non-axisymmetric) perturbations if $\hat{\chi}^{1}<\hat{\chi}^{0}$ ($\hat{\chi}^{1}>\hat{\chi}^{0}$) when $\hat{\chi}<$max$\{\hat{\chi}^{0},\hat{\chi}^{1}\}$. Moreover, fixed contact lines can be represented as the limiting case of free contact lines when $\chi \rightarrow \infty$. Therefore, for a fixed $\Gamma_{gl}$, $\lambda \rightarrow -\infty$ as $\chi \rightarrow -\infty$ and $\lambda \rightarrow \nu$ as $\chi \rightarrow \infty$; here, $\lambda$ is the smallest eigenvalue of the Sturm-Liouville problem, and $\nu$ is the smallest eigenvalue of a similar problem with $\varphi_{i}=0$ at $\ell$. Hence,
\begin{equation}
	\lambda \leq \nu,
	 \label{eqn:eq24}
\end{equation}
implying that the stability region of capillary surfaces with free contact lines is a subset of the corresponding stability region for the same capillary surfaces with fixed contact lines\footnote{What `the same' means here depends on how a capillary surface is specified. For example, as will be discussed in section \ref{sec:equilibrium}, a catenoid, such as the one shown in Fig.~\ref{fig:figure1}, is uniquely specified by $\hat{s}_{0}$ and $\hat{s}_{1}$. Hence, catenoids with free contact lines are being compared to those with the same $\hat{s}_{0}$ and $\hat{s}_{1}$, but fixed at $\hat{s}_{1}$.}. The latter is termed the maximal stability region (MSR), {\color{black}a concept introduced by \citet{slobozhanin1974characteristic}.} The critical states are determined by
\begin{equation}
	D^{0}=
    	\left| \begin{array} {c c c}
	w_{1}(\hat{s}_{0}) & w_{2}(\hat{s}_{0}) & w_{3}(\hat{s}_{0})\\
	w_{1}(\hat{s}_{1}) & w_{2}(\hat{s}_{1}) & w_{3}(\hat{s}_{1})\\
	\int_{\hat{s}_{0}}^{\hat{s}_{1}} \hat{r} w_{1}\mbox{d}\hat{s} & \int_{\hat{s}_{0}}^{\hat{s}_{1}} \hat{r} w_{2}\mbox{d}\hat{s} & \int_{\hat{s}_{0}}^{\hat{s}_{1}} \hat{r} w_{3}\mbox{d}\hat{s}
	\end{array} \right|,
	 \label{eqn:eq25}
\end{equation}

\begin{equation}
	D^{1}=
    	\left| \begin{array} {c c }
	w_{4}(\hat{s}_{0}) & w_{5}(\hat{s}_{0})\\
	w_{4}(\hat{s}_{1}) & w_{5}(\hat{s}_{1})
	\end{array} \right|.
	 \label{eqn:eq26}
\end{equation}
For a fixed $\hat{s}_{0}$, the first $\hat{s}_{1}$ along the meridian curve at which $D^{0}=0$ ($D^{1}=0$) corresponds to a critical state of the MSR with respect to axisymmetric (non-axisymmetric) perturbations. Note that the MSR coincides with the stability region for capillary surfaces with only pinned contact lines. Moreover, determining the MSR for capillary surfaces with free contact lines prior to testing the stability criteria given by Eqs.~(\ref{eqn:eq22}) and (\ref{eqn:eq23}) is necessary, since $\hat{\chi}=\hat{\chi}^{0}$ and $\hat{\chi}=\hat{\chi}^{1}$ generally have more than one solution. Therefore, $\hat{\chi}>$max$\{\hat{\chi}^{0},\hat{\chi}^{1}\}$ indicates stability only for surfaces belonging to the MSR. All equilibrium surfaces outside the MSR are unstable. {\color{black}A summary of Myshkis's method is given by \citet{bostwick2015stability}.}

\section{Results and discussion} \label{sec:result}
\subsection{Equilibrium solution} \label{sec:equilibrium}

Solving Eq.~(\ref{eqn:eq3}) for $q=0$ furnishes the equilibrium meridian curve
\begin{equation}
	\left\{ \begin{array}{l}
         		\hat{r}(\hat{s})=\sqrt{\hat{s}^2+1}\\
		\hat{z}(\hat{s})=-\ln(\hat{s}+\sqrt{\hat{s}^2+1})\\
	\end{array} \right.
    	 \label{eqn:eq27}
\end{equation}
with
\begin{equation}
	\Lambda=\frac{1}{\sqrt{\hat{s}_{0}^2+1}}\ln\left(\frac{\hat{s}_{1}+\sqrt{\hat{s}_{1}^2+1}}{\hat{s}_{0}+\sqrt{\hat{s}_{0}^2+1}} \right),
    	 \label{eqn:eq28}
\end{equation}

\begin{equation}
	V=\frac{\hat{s}_{1}\sqrt{\hat{s}_{1}^2+1}-\hat{s}_{0}\sqrt{\hat{s}_{0}^2+1}+\Lambda\sqrt{\hat{s}_{0}^2+1}}{2\Lambda(\hat{s}_{0}^2+1)^{3/2}},
    	 \label{eqn:eq29}
\end{equation}

\begin{equation}
	\theta_{d}=\tan^{-1}(1/\hat{s}_{0}),
    	 \label{eqn:eq30}
\end{equation}

\begin{equation}
	\theta_{c}=\tan^{-1}(-1/\hat{s}_{1}).
    	 \label{eqn:eq31}
\end{equation}
Equations (\ref{eqn:eq28})-(\ref{eqn:eq31}) furnish four constraints on $\Lambda$, $V$, $\theta_{c}$, $\theta_{d}$, $\hat{s}_{0}$, and $\hat{s}_{1}$, leaving two degrees of freedom. Fixing two variables, the other four and the catenoid geometry, as shown in Fig.~\ref{fig:figure1}, are completely specified. Therefore, among these six, one can select two variables to represent stability and equilibrium data as two-dimensional phase diagrams. Furthermore, Eqs.~(\ref{eqn:eq30}) and (\ref{eqn:eq31}) provide a one-to-one correspondence between $(\hat{s}_{0},\hat{s}_{1})$ and $(\theta_{c},\theta_{d})$. Thus, they can be interchanged without affecting phase diagram characteristics, albeit $(\theta_{c},\theta_{d})$ are preferred to $(\hat{s}_{0},\hat{s}_{1})$ as they vary in a finite range. In this paper, the catenoid equilibrium solution and stability region are presented with respect to $\Lambda$, $V$, $\theta_{c}$, and $\theta_{d}$ subject to various constraints.

The cylindrical volume and slenderness are the two favourable quantities with which stability regions have been represented in the literature \citep{martinez1986liquid, myshkis1987low, bayramli1987experimental, slobozhanin1993stability, lowry1995capillary, slobozhanin1996stability} because they can be readily measured experimentally. We refer to $(\Lambda, V)$ as `favourable parameters' and the respective phase diagram as the `favourable phase diagram'. Moreover, Eqs.~(\ref{eqn:eq28})-(\ref{eqn:eq31}) are single-valued functions of $\hat{s}_{0}$ and $\hat{s}_{1}$. Consequently, the left-hand-side parameters characterizing a catenoid described in Fig.~\ref{fig:figure1} are uniquely specified with respect to $\hat{s}_{0}$ and $\hat{s}_{1}$. Thus, $(\hat{s}_{0},\hat{s}_{1})$ are more convenient for representing the MSR and stability region. We refer to $(\hat{s}_{0},\hat{s}_{1})$ or $(\theta_{c},\theta_{d})$, as `canonical parameters' and the respective phase diagram as the `canonical phase diagram'.

 \begin{figure} [t]
	\centering
	\includegraphics[width=\linewidth]{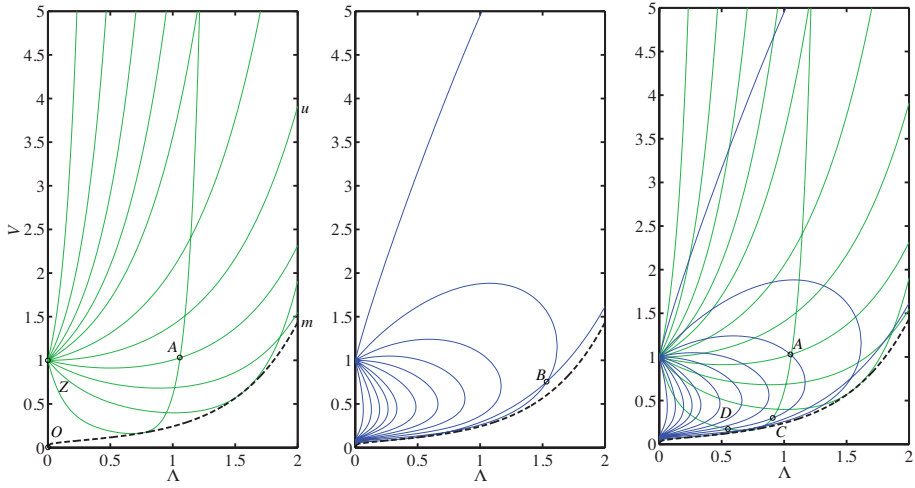}
 	\caption{Equilibrium solution and existence region (above dashed line) with respect to the favourable parameters; (a) constant-dihedral angle isocontours (labels denote $\theta_{d}$ in degrees), (b) constant-contact angle isocontours (labels denote $\theta_{c}$ in degrees), and (c) constant-dihedral and contact angle isocontours.}
	\label{fig:figure2}
\end{figure}

Identifying the existence-region boundary is necessary to properly represent equilibrium solutions with respect to the favourable and canonical parameters. This is straightforward for the canonical parameters because $\theta_{d}\in(\pi-\theta_{c},\pi]$ for $\theta_{c} \in [0,\pi]$. However, determining the existence-region boundary with respect to the favourable parameters is non-trivial. One may naturally suppose that $V(\hat{s}_{0},\hat{s}_{1},\Lambda)$ has a minimum for a given $\Lambda$. Therefore, the cylindrical volume given by Eq.~(\ref{eqn:eq29}) is to be minimized subject to a constant slenderness $\Lambda$, leading to
 \begin{equation}
	\left(\frac{\partial V}{\partial \hat{s}_{0}} \right)_{\hat{s}_{1},\Lambda}+\vartheta \left(\frac{\partial \Lambda}{\partial\hat{s}_{0}} \right)=0,
    	 \label{eqn:eq32}
\end{equation}
 \begin{equation}
	\left(\frac{\partial V}{\partial \hat{s}_{1}} \right)_{\hat{s}_{0},\Lambda}+\vartheta \left(\frac{\partial \Lambda}{\partial\hat{s}_{1}} \right)=0,
    	 \label{eqn:eq33}
\end{equation}
\begin{equation}
	\Lambda(\hat{s}_{0},\hat{s}_{1})=\mbox{const.},
    	 \label{eqn:eq34}
\end{equation}
where $\vartheta$ is a Lagrange multiplier. Moreover, Eqs.~(\ref{eqn:eq32})-(\ref{eqn:eq34}) define the existence-region boundary in the $\Lambda$-$V$ space, also furnishing a lower bound on $V$.

Figure~\ref{fig:figure2} shows equilibrium isocontours and existence region with respect to the favourable parameters. The $Om$ curve (dashed line) is the existence-region boundary that corresponds to $(\Lambda,V)$ satisfying Eqs.~(\ref{eqn:eq32})-(\ref{eqn:eq34}). No catenoid can be found with volume and slenderness below this curve. 

Iso-$\theta_{d}$ curves are plotted in Fig.~\ref{fig:figure2}(a), facilitating the representation of solution multiplicity with respect to $\Lambda$, $V$, and $\theta_{d}$. Isocontours can be viewed as the level curves of a multivalued function $\theta_{d}=\theta_{d}(\Lambda,V)$. For example, the level curves with $\theta_{d}=110^{\circ}$ and $\theta_{d}=170^{\circ}$ intersect at $A$, implying that two equilibrium solutions exist for the corresponding $(\Lambda,V)$. The isocontour with $\theta_{d}=\pi/2$ is a special case that asymptotes to $Om$. It can be proved that
\begin{equation}
 	\lim_{\Lambda \to \infty} V(\hat{s}_{0},\hat{s}_{1})|_{(\hat{s}_{0},\hat{s}_{1}) \in Zu}=\lim_{\Lambda \to \infty} V(\hat{s}_{0},\hat{s}_{1}) |_{(\hat{s}_{0},\hat{s}_{1}) \in Om} =\lim_{\Lambda \to \infty} \frac{\sinh^{2} \Lambda}{2 \Lambda}=\infty,
    	 \label{eqn:eq35}
\end{equation}
where $Zu$ is the isocontour with $\theta_{d}=\pi/2$. Figure~\ref{fig:figure2}(b) can be similarly interpreted. Here, isocontours can be viewed as the level curves of a multivalued function $\theta_{c}=\theta_{c}(\Lambda,V)$. Isocontours with $\theta_{c}=150^{\circ}$ and $\theta_{c}=170^{\circ}$ intersect at $B$, implying that two equilibrium solutions exist for the corresponding $(\Lambda,V)$. For a given $\theta_{c}$, there is a slenderness above which there are no catenoids. Two equilibrium solutions exist for smaller slendernesses. Iso-$\theta_{d}$ and iso-$\theta_{c}$ curves are plotted together in Fig.~\ref{fig:figure2}(c). This figure demonstrates that isocontour intersections do not always correspond to equilibrium solutions. For example, the isocontours with $\theta_{d}=170^{\circ}$ and $\theta_{c}=130^{\circ}$ intersect at $A$, $C$, and $D$ corresponding to three different $(\Lambda,V)$. However, $D$ is the only intersection representing an equilibrium at $\theta_{d}=170^{\circ}$ and $\theta_{c}=130^{\circ}$. Note that these isocontours are projections of the corresponding level curves onto the $(\Lambda,V)$ plane. Here, $A$ and $C$ lie at the intersections of the projections and do not represent equilibrium solutions. Hence, the space $(\Lambda,V)$ is unsuitable for representing the stability of catenoids because it cannot be partitioned into mutually exclusive regions of stability and instability.

\begin{figure} 
	\centering
	\includegraphics[width=\linewidth]{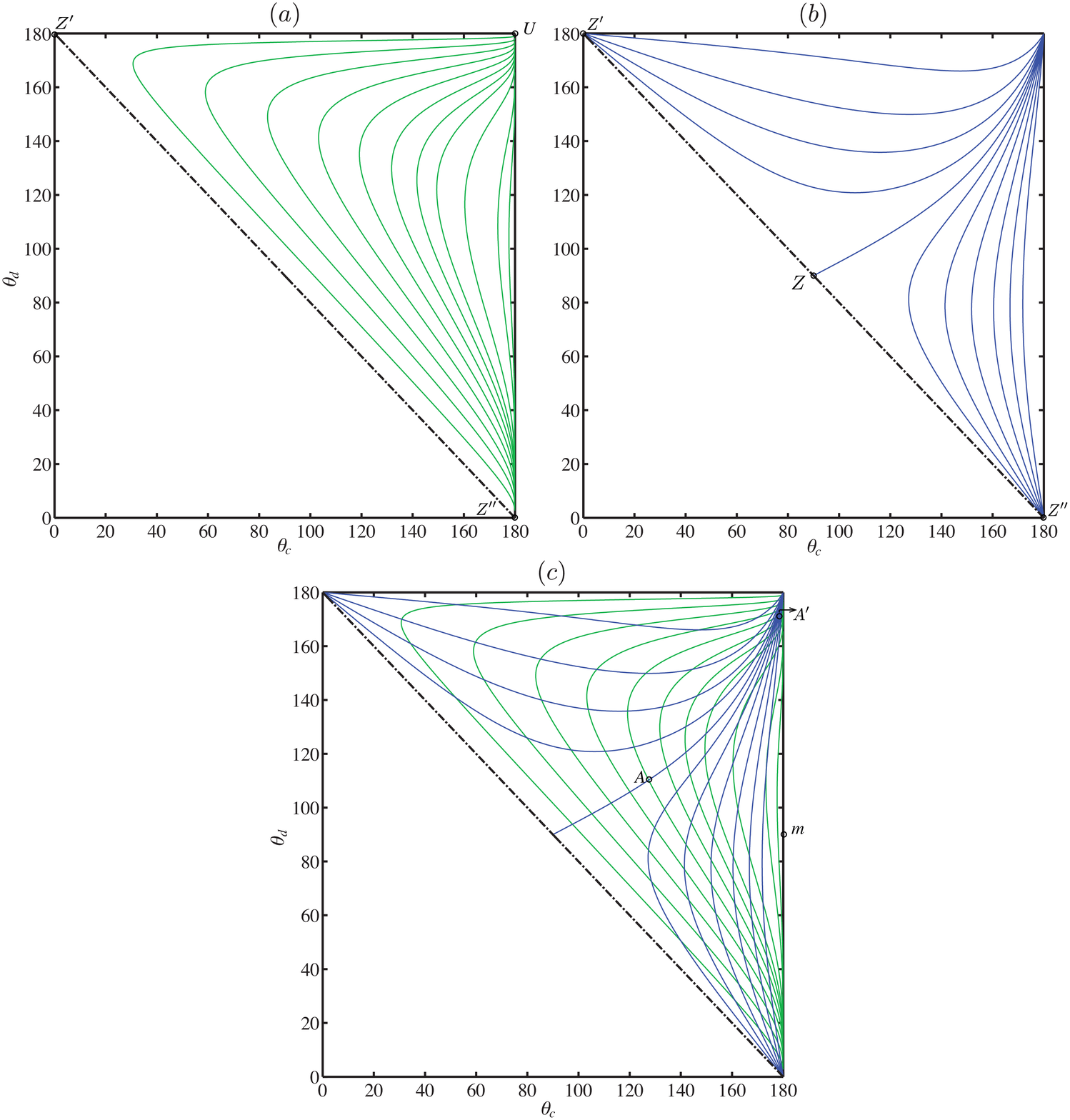}	
 	\caption{Equilibrium solution and existence region (above dash-dotted line) with respect to the canonical parameters; (a) constant-slenderness isocontours (label denote $\Lambda$), (b) constant-cylindrical volume isocontours (labels denote $V$), and (c) constant-slenderness and cylindrical volume isocontours.}
	\label{fig:figure3}
 \end{figure}

Figure~\ref{fig:figure3} shows equilibrium isocontours and the existence region with respect to the canonical parameters. As previously discussed, equilibrium solutions for catenoids cannot be conveniently represented with respect to the favourable parameters because isocontour intersections are not always associated with equilibrium solutions. Moreover, two solutions may correspond to the same point $(\Lambda,V)$ in the existence region. In contrast, the canonical parameters furnish a one-to-one correspondence between points in the existence region and equilibrium solutions. Here, the existence region is the upper triangle indicated by $Z'Z''U$. Note that the existence-region boundary $Z'Z''$ does not correspond to $Om$ in Fig.~\ref{fig:figure2}. Iso-$\Lambda$ and iso-$V$ curves are plotted in Figs.~\ref{fig:figure3}(a) and (b). Isocontours are the level curves of the single-valued functions $\Lambda=\Lambda(\theta_{c},\theta_{d})$ and $V=V(\theta_{c},\theta_{d})$ given by Eqs.~(\ref{eqn:eq28}) and (\ref{eqn:eq29}). Iso-$\Lambda$ and iso-$V$ curves are overlaid in Fig.~\ref{fig:figure3}(c). Here, unlike the favourable parameters, every isocontour intersection uniquely represents an equilibrium solution. Interesting to note are the two equilibrium solutions corresponding to the point $A$ in Fig.~\ref{fig:figure2}, which are denoted by $A$ and $A'$, and are distinctly represented with respect to the canonical parameters.
 
\subsection{Stability} \label{sec:stability}
Several factors affect the equilibrium state and stability of capillary surfaces, including fluid inertia, external fields (\eg, gravitational and centrifugal forces), and boundary conditions at contact lines. The latter differentiates contact-drop dispensing applications from classical liquid bridge problems where the equilibrium surface is pinned to two coaxial discs. The contact-line condition can be easily accounted for in the equilibrium solution by integration constants of the integral curve obtained from Eq.~(\ref{eqn:eq3}). The influence on stability, however, is not straightforward. It affects the eigenvalues of the Sturm-Liouville problem through the boundary condition at $\ell$. This plays a far more significant role in the stability of capillary surfaces. For example, the notion of wavenumber introduced for classifying equilibrium solution branches and characterizing the bifurcation of liquid bridges \citep{lowry1995capillary, slobozhanin2002stability} is directly related to the conditions at the contact lines. Before proceeding to catenoids, we elucidate the contact-line condition effect on the stability limit of cylindrical liquid bridges. 

 \begin{figure} 
\centering
\includegraphics[width=0.7\linewidth]{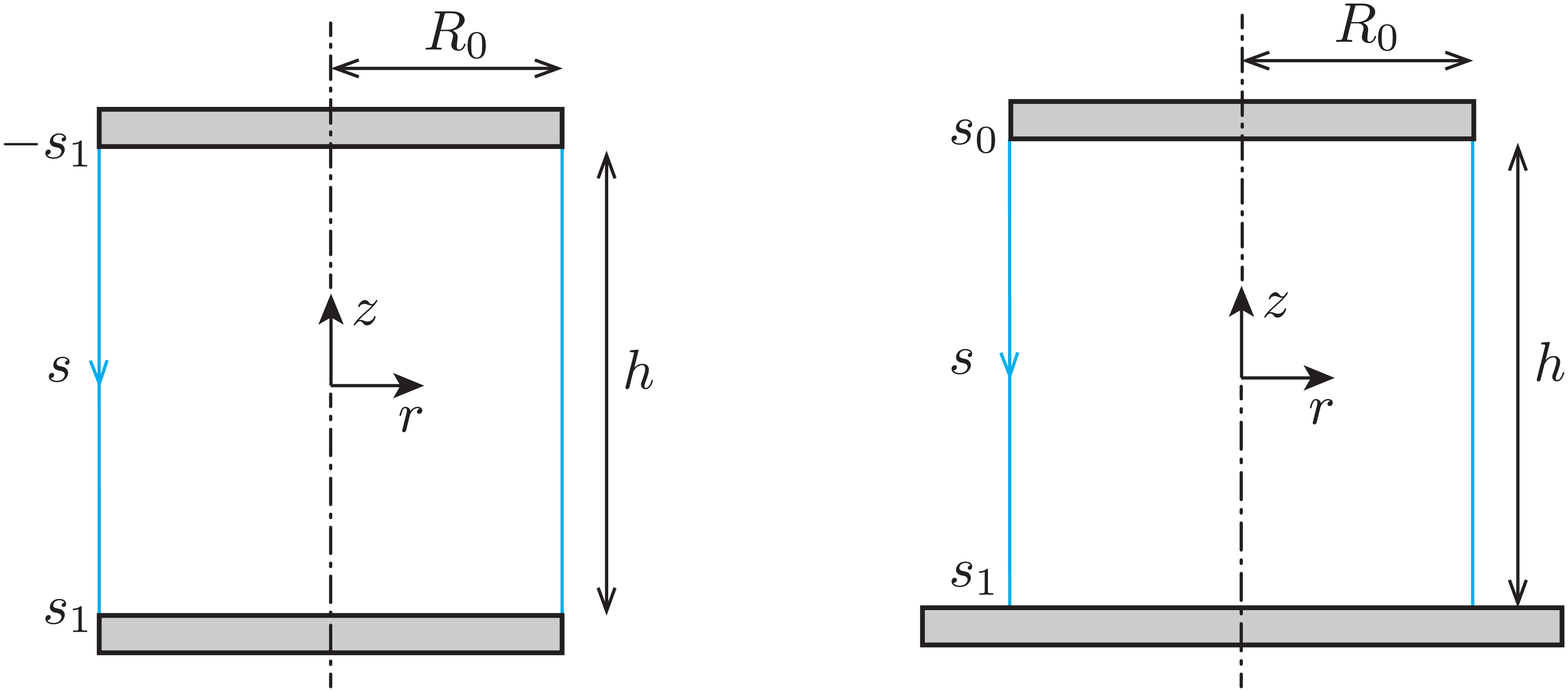}
 \caption{Cylindrical liquid bridge: equilibrium surface is pinned to both discs (left); equilibrium surface is pinned at the upper disc and free to move on the lower plate (right).}
 \label{fig:figure4}
 \end{figure}

\subsubsection{Cylinder} \label{sec:stability:cylinder}

 {\color{black} \citet{johns2002interfacial} determined the static stability limit of cylindrical liquid bridges with a pinned and a free contact line using perturbation techniques. Here, we derive the stability criteria using Myshkis's variational method.} Consider the cylindrical liquid bridge between two plates shown in Fig.~\ref{fig:figure4}. \citet{plateau1873statique} theoretically obtained the stability region $\Lambda<2\pi$ for cylindrical liquid bridges pinned at two equal coaxial discs (Fig.~\ref{fig:figure4} (left)). In this section, the corresponding stability limit is obtained for cylindrical liquid bridges that are pinned to a disc and free to move on a plate (Fig.~\ref{fig:figure4} (right)). All the lengths are scaled with $q$ as
\begin{equation}
	\rho=|q|r, \quad  \xi=qz, \quad \tau=|q|s
    	 \label{eqn:eq36}
\end{equation}
with
\begin{equation}
	L \equiv \frac{\mathcal{L}}{\hat{q}^2} = \frac{\mbox{d}^{2}}{\mbox{d}\tau^{2}}+\frac{\rho'}{\rho}\frac{\mbox{d}}{\mbox{d}\tau}+\left[ \left( 1-\frac{\xi'}{\rho}\right)^{2}+\left(\frac{\xi'}{\rho}\right)^{2}\right]
    	 \label{eqn:eq37}
\end{equation}
used in Eqs.~(\ref{eqn:eq17})-(\ref{eqn:eq21}) instead of $\mathcal{L}$. The solutions of Eqs.~(\ref{eqn:eq17})-(\ref{eqn:eq21}) are
\begin{equation}
	w_{1}(\tau)=\sin \tau, \quad w_{2}(\tau)=\cos \tau, \quad w_{3}(\tau)=-1
    	 \label{eqn:eq38}
\end{equation}
for axisymmetric and
\begin{equation}
	w_{4}(\tau)=\tau, \quad w_{5}(\tau)=1
    	 \label{eqn:eq39}
\end{equation}
for non-axisymmetric perturbations. These furnish 
\begin{equation}
	D^{0}(\Delta\tau)=-\Delta\tau\sin \Delta\tau+2(1-\cos \Delta\tau),
    	 \label{eqn:eq40}
\end{equation}
\begin{equation}
	D^{1}(\Delta\tau)=\Delta\tau,
    	 \label{eqn:eq41}
\end{equation}
\begin{equation}
	\tilde{\chi}^{0}(\Delta\tau)=\frac{\Delta\tau\cos \Delta\tau-\sin \Delta\tau}{-\Delta\tau\sin \Delta\tau+2(1-\cos \Delta\tau)},
    	 \label{eqn:eq42}
\end{equation}
\begin{equation}
	\tilde{\chi}^{1}(\Delta\tau)=-\frac{1}{\Delta\tau},
    	 \label{eqn:eq43}
\end{equation}
where $\Delta\tau=\tau_{1}-\tau_{0}$ and $\tilde{\chi}=\chi/|q|$. The MSR boundary is determined by $D^{0}(\Delta\tau)=0$ and $D^{1}(\Delta\tau)=0$ with respect to axisymmetric and non-axisymmetric perturbations, respectively. The first root of Eq.~(\ref{eqn:eq40}) occurs where $\Delta\tau^{0}_{MSR}=2\pi$, whereas Eq.~(\ref{eqn:eq41}) has no non-trivial root. This implies that all cylindrical bridges with $\Delta\tau>2\pi$ are unstable to axisymmetric perturbations, irrespective of the contact-line condition at $s_{1}$. Note that the MSR of Fig.~\ref{fig:figure4} (right) is equivalent to the stability region of Fig.~\ref{fig:figure4} (left), and, therefore, the foregoing condition coincides with Plateau's stability criterion. The first non-trivial root of $\tilde{\chi}^{0}=\tilde{\chi}$ and $\tilde{\chi}^{1}=\tilde{\chi}$ inside the respective MSR identifies the stability region with respect to axisymmetric and non-axisymmetric perturbations, respectively. The former gives $\Delta\tau^{0}_{cr} \simeq 4.4934$, whereas the latter has no non-trivial root. For cylindrical liquid bridges, $\Lambda$ and $\Delta\tau$ are equal; thus, the MSR and stability region can be summarized as $\Lambda<2\pi$ and $\Lambda<4.4934$ where axisymmetric perturbations are the most dangerous. Note that the stability region of cylindrical liquid bridges with two free contact lines is $\Lambda<\pi$ \citep{langbein2002capillary}, which clearly indicates the destabilizing effect of free contact lines.

\subsubsection{Catenoid} \label{sec:stability:catenoid}
Here, we apply the same procedure as for cylindrical liquid bridges. Solving Eqs.~(\ref{eqn:eq17})-(\ref{eqn:eq21}) using the integral curve of Eq.~(\ref{eqn:eq27}) gives
\begin{equation}
	w_{1}(\hat{s})=\frac{\hat{s}}{\sqrt{\hat{s}^2+1}},
    	 \label{eqn:eq44}
\end{equation}
\begin{equation}
	w_{2}(\hat{s})=1-\frac{\hat{s}}{\sqrt{\hat{s}^2+1}}\ln(\hat{s}+\sqrt{\hat{s}^2+1}),
    	 \label{eqn:eq45}
\end{equation}
\begin{equation}
	w_{3}(\hat{s})=-\frac{\hat{s}^2+1}{4},
    	 \label{eqn:eq46}
\end{equation}
\begin{equation}
	w_{4}(\hat{s})=\frac{\hat{s}}{2}+\frac{\hat{s}}{2\sqrt{\hat{s}^2+1}}\ln(\hat{s}+\sqrt{\hat{s}^2+1}),
    	 \label{eqn:eq47}
\end{equation}
\begin{equation}
	w_{5}(\hat{s})=\frac{1}{\sqrt{\hat{s}^2+1}}.
    	 \label{eqn:eq48}
\end{equation}
Substituting Eqs.~(\ref{eqn:eq44})-(\ref{eqn:eq48}) into Eqs.~(\ref{eqn:eq22}), (\ref{eqn:eq23}), (\ref{eqn:eq25}), and (\ref{eqn:eq26}) furnishes $D^{0}$, $D^{1}$, $\hat{\chi}^{0}$, and $\hat{\chi}^{1}$ as functions of $\hat{s}_{0}$ and $\hat{s}_{1}$. These, unlike for cylinders, cannot generally be represented as functions of only $\Delta\hat{s}$, implying that the stability of catenoids demands two independent parameters to be completely specified. This is consistent with the equilibrium solution discussed in section~\ref{sec:equilibrium}.
 \begin{figure} 
	\centering
	\includegraphics[width=\linewidth]{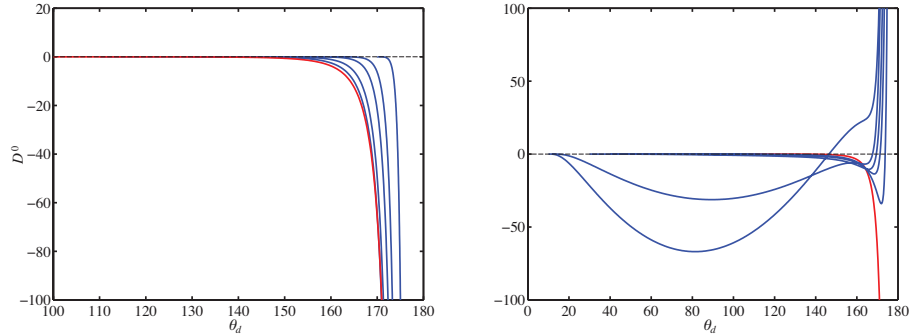}	
 	\caption{The effect of contact angle $\theta_{c}$ on $D^{0}$; left panel: $\theta_{c}=10, 20, 30,50, 70^{\circ}$ (blue, right to left) and $\theta_{c}=90^{\circ}$ (red); right panel: $\theta_{c}=130,140,150,168,170^{\circ}$ (blue, right to left) and $\theta_{c}=90^{\circ}$ (red).}
	\label{fig:figure5}
 \end{figure}
 
\citet{erle1970stability} showed that catenoids pinned to two equal coaxial discs are unstable to axisymmetric perturbations when $\Delta\hat{s}/2>4.6395$. We will determine the stability region for catenoids with a free contact line as shown in Fig.~\ref{fig:figure1} and demonstrate that they lose stability to axisymmetric perturbations. This can be accomplished by showing that $D^{1}=0$ and $\hat{\chi}^{1}=\hat{\chi}$ have no non-trivial root (proved in Appendix~\ref{sec:appendixa}). Accordingly, the MSR and stability region with respect to non-axisymmetric perturbations coincide with the existence region. Moreover, one can prove that $D^{0}=0$ has a non-trivial root only when $\theta_{c}>\pi/2$ (Appendix~\ref{sec:appendixa}). This is clearly illustrated in Fig.~\ref{fig:figure5}. Here, $D^{0}(\theta_{c},\theta_{d})$ is plotted in Fig.~\ref{fig:figure5} (left) for $\theta_{c}\leq \pi/2$ and $\theta_{d} \in (\pi-\theta_{c},\pi]$. For a given $\theta_{c}$, $D^{0} \to 0^{-}$ as $\theta_{d} \to \pi-\theta_{c}$ and $D^{0} \to -\infty$ as $\theta_{d} \to \pi$. Thus, no $\hat{s}_{0}$ can be found along the integral curve where $D^{0}$ vanishes, and the MSR spans the entire existence region. In contrast, for a given $\theta_{c}>\pi/2$, there exists a $\theta_{d}$ (or $\hat{s}_{0}$) at which $D^{0}$ vanishes, as indicated in Fig.~\ref{fig:figure5} (right). Here, $D^{0} \to 0^{-}$ as $\theta_{d} \to \pi-\theta_{c}$ and $D^{0} \to \infty$ as $\theta_{d} \to \pi$. Note that $\theta_{c}\simeq168.75^{\circ}$ is a special case because $D^{0}$ and $\partial D^{0}/\partial \theta_{d}$ vanish simultaneously at $\theta_{d}\simeq162.07^{\circ}$.

\begin{figure}
	\centering
	\includegraphics[width=\linewidth]{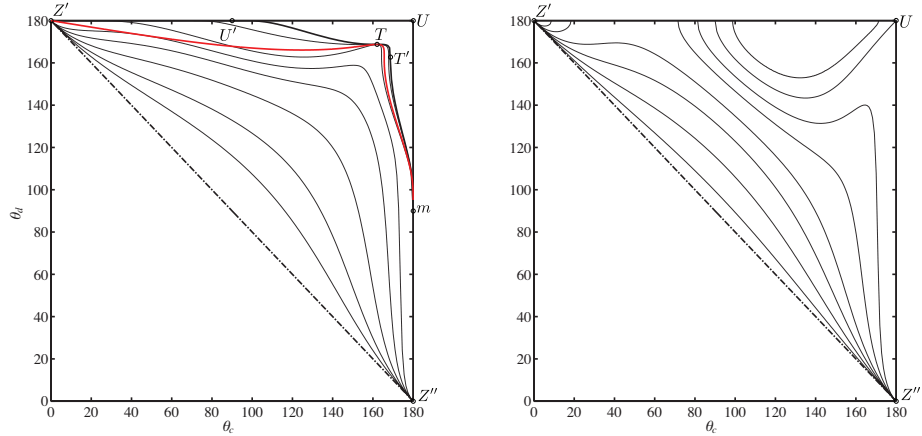}	
 	\caption{Canonical phase diagram; (a) axisymmetric perturbations: constant-$\hat{\chi}^{0}$ isocontours (thin-solid lines, labels denote $\hat{\chi}^{0}$), vanishingly small catenoids as $\hat{\chi}^{0}\to-\infty$ (thick-dash-dotted line), the MSR boundary as $\hat{\chi}^{0}\to\infty$ (thick-solid line), and the stability region boundary (dashed line); (b) non-axisymmetric perturbations: constant-$\hat{\chi}^{1}$ isocontours (thin-solid lines, labels denote $\hat{\chi}^{1}$), and vanishingly small catenoids as $\hat{\chi}^{1}\to-\infty$ (thick-dash-dotted line).}
	\label{fig:figure6}
\end{figure}

Figure~\ref{fig:figure6}(a) shows the canonical phase diagram representing the MSR and stability region for axisymmetric perturbations. The regions confined by $Z'Z''mT'TU'Z'$ and $Z'Z''mTZ'$ represent the MSR and stability region, respectively. The MSR boundary $mT'TU'$ and stability-region boundary $mTZ'$ are determined, respectively, by $D^{0}(\hat{s}_{0},\hat{s}_{1})=0$ and $\hat{\chi}^{0}(\hat{s}_{0},\hat{s}_{1})=\hat{\chi}(\hat{s}_{1})$. The meridian curve for catenoids corresponding to points on the MSR boundary satisfies $\nu=0$, and that corresponding to points on the stability region boundary satisfies $\lambda=0$. All the catenoids corresponding to points outside the MSR $mUU'TT'm$ are, regardless of the contact-line condition at $\hat{s}_{1}$, unstable to axisymmetric perturbations. $D^{0}$ and $\partial D^{0}/\partial \theta_{c}$ vanish simultaneously at $T$ where $(\theta_{c},\theta_{d})\simeq(162.07^{\circ},168.75^{\circ})$. Similarly, $D^{0}$ and $\partial D^{0}/\partial \theta_{d}$ vanish simultaneously at $T'$ where $(\theta_{c},\theta_{d})\simeq(168.75^{\circ},162.07^{\circ})$. Note that the MSR here is equivalent to the stability region of catenoids pinned to two unequal coaxial discs. Hence, Fig.~\ref{fig:figure6}(a) also allows a comparison between two stability problems: (1) Catenoids pinned to a disc and free to move on a plate (Fig.~\ref{fig:figure1}), and (2) catenoids pinned to two unequal coaxial discs with exactly the same $\hat{s}_{0}$ and $\hat{s}_{1}$. The region confined by $mT'TU'Z'Tm$ represents the catenoids that are unstable to axisymmetric perturbations in the first problem, but stable in the second. Iso-$\hat{\chi}^{0}$ curves are thin solid black lines approaching $Z'Z''$ ($mT'TU'$) as $\hat{\chi}^{0} \to -\infty$ ($\hat{\chi}^{0} \to \infty$). Therefore, catenoids corresponding to points in the close vicinity of $Z'Z''$ ($mT'TU'$) are highly stable (unstable) since $\hat{\chi}_{0}\ll\hat{\chi}$ ($\hat{\chi}_{0}\gg\hat{\chi}$). Figure~\ref{fig:figure6}(b) shows the canonical phase diagram representing the MSR and stability region for non-axisymmetric perturbations. Here, $D^{1}(\hat{s}_{0},\hat{s}_{1})=0$ and $\hat{\chi}^{1}(\hat{s}_{0},\hat{s}_{1})=\hat{\chi}(\hat{s}_{1})$ have no non-trivial solution. Thus, the MSR and stability region coincide with the existence region, implying that catenoids are always stable with respect to non-axisymmetric perturbations. Iso-$\hat{\chi}^{1}$ curves are plotted as thin solid black lines approaching $Z'Z''$ as $\hat{\chi}^{1} \to -\infty$. Note that $\hat{\chi}^{1}$ does not approach infinity for isocontours near the existence-region boundary $Z'UZ''$.
 
Figure~\ref{fig:figure6} also illustrates how the catenoid geometrical symmetry is reflected in its phase diagram. Catenoids that are pinned to two equal coaxial discs \citep{erle1970stability} have equatorial symmetry, resulting in a one-dimensional phase digram in $\Delta\hat{s}$. Even though catenoids bridging two unequal coaxial discs generally have no equatorial symmetry, and they require a two-dimensional phase diagram, a symmetric stability region can be constructed by choosing a proper set of parameters. For instance, one may choose the ratio of the lower and upper disc diameters $K$ to represent the phase diagram (the second parameter can arbitrarily be selected). These catenoids are reflectively symmetric with respect to $K$. Clearly, inverting this ratio has no effect on the stability limit. Hence, the stability-region boundary must be invariant with respect to the transformation $K=1/\bar{K}$. Alternatively, one can choose the dihedral angle that the catenoid forms with the upper disc $\theta_{d}$ and the lower one $\theta_{c}$. The foregoing transformation can equivalently be written
\begin{equation}
	\left\{ \begin{array}{l}
         		\theta_{c}=\bar{\theta}_{d}\\
		\theta_{d}=\bar{\theta}_{c}\\
	\end{array} \right. ,
    	 \label{eqn:eq49}
\end{equation}
which is why the MSR boundary in Fig.~\ref{fig:figure6} (left) is symmetric with respect to the phase diagram minor diagonal described by $\theta_{d}=\theta_{c}$. This is formally proved in Appendix~\ref{sec:appendixb}. Note that the stability-region boundary has no such symmetry since the contact line condition at $\ell$ (see Fig.~\ref{fig:figure1}) completely breaks the equatorial and reflective symmetries.

\begin{figure}
	\centering
	\includegraphics[width=\linewidth]{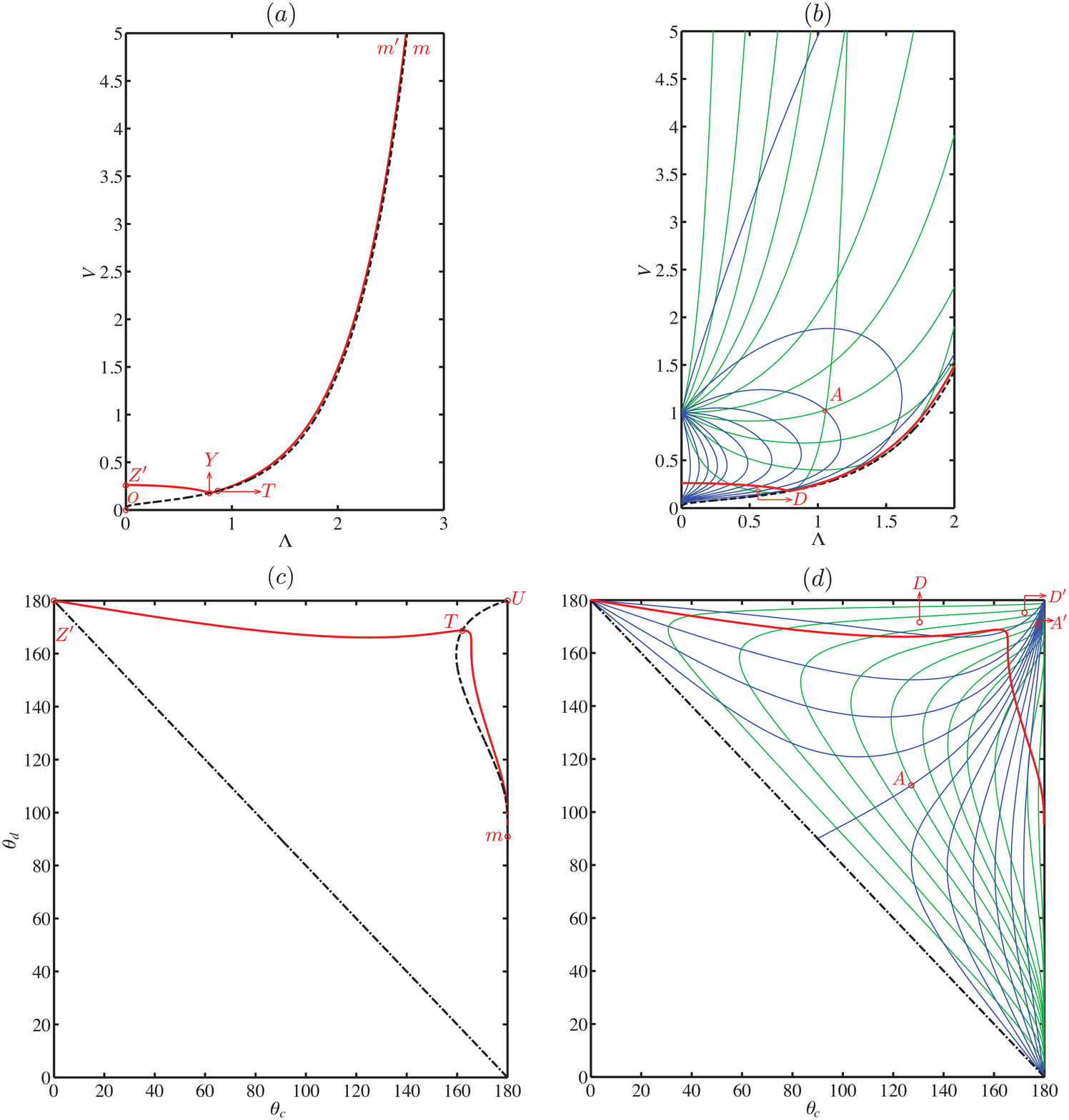}	
 	\caption{Existence region in the canonical phase diagram, existence-region boundary (dash-dotted for canonical and thin-short-dashed for favourable), stability-region boundary with respect to volume-controlled (thick-short-dashed) and pressure-controlled (long-dashed) perturbations; (a) stability region in the canonical phase diagram with (b) the respective isocontours (the same as Fig.~\ref{fig:figure3}).}
	\label{fig:figure7}
\end{figure}

{\color{black}
Figure~\ref{fig:figure7}(a) shows the existence-region boundary $UTm$ in the favourable phase diagram ($Om$ in Fig.~\ref{fig:figure2}) and stability-region boundary $Z'Tm$. The curve $UTm$ is the locus of points at which an iso-$\theta_{d}$ curve is tangent to an iso-$\theta_{c}$ curve. The curve corresponding to the existence-region boundary with respect to the favourable parameters $UTm$ intersects the stability region boundary at $T$ where $(\theta_{c},\theta_{d})\simeq(162.07^{\circ},168.75^{\circ})$. This is the point at which the slope of the MSR boundary is zero, as discussed for Fig.~\ref{fig:figure6}(a). Figure~\ref{fig:figure7}(a) clearly demonstrates that all the points on the segments $UT$ and $Tm$ correspond to unstable and stable catenoids, respectively. Figure~\ref{fig:figure7}(b) shows isocontours in the canonical phase diagrams. Selecting two variables among $\Lambda$, $V$, $\theta_{c}$, and $\theta_{d}$, this figure describes the equilibrium solution and stability of the corresponding catenoids. Consider the point $(\Lambda,V)\simeq(1.0518,1.0264)$ in the favourable diagram, for example. It lies at the intersection of $\theta_{c}=130^{\circ}$ and $\theta_{d}=110^{\circ}$. This can be located in the canonical phase diagram, as shown in Fig.~\ref{fig:figure7}(b). The corresponding point $A$ is inside the stability region, indicating that the respective catenoid is stable. Furthermore, the second equilibrium solution can be determined by identifying the other intersection point of the same iso-$\Lambda$ and iso-$V$. This occurs at $A'$, where $(\theta_{c},\theta_{d})\simeq(177.08^{\circ},170.1^{\circ})$. The second solution lies outside the stability region, which corresponds to an unstable catenoid. Consider the point $(\Lambda,V)\simeq(0.5555,0.1742)$ in the favourable diagram. The two equilibrium solutions are represented in Fig.~\ref{fig:figure7}(b) by $D$ at $(\theta_{c},\theta_{d})=(130^{\circ},170^{\circ})$ and $D'$ at $(\theta_{c},\theta_{d})\simeq(175.32^{\circ},174.96^{\circ})$. Both equilibrium solutions lie outside the stability region and correspond to unstable catenoids.

Volume-controlled catenoids (constrained) are exposed to a smaller set of disturbances than pressure-controlled catenoids (unconstrained). Therefore, the former is more constrained and stable. Using Maddocks' theorems \citep{maddocks1987stability}, \citet{akbari2014bridge} showed that catenoids lose stability with respect to constant-pressure perturbations at turning points in $\Lambda$ along iso-$\theta_{c}$ curves (see Fig.~\ref{fig:figure2}(b)). Accordingly, the upper and lower segments of these isocontours respectively correspond to stable and unstable catenoids. Figure~\ref{fig:figure7}(b) compares constant-volume and constant-pressure stability regions. Here, the region confined between the long-dashed and dashed-dotted lines is the constant-pressure stability region and that between the thick-short-dashed and dashed-dotted lines is the constant-volume stability region. As expected, the constant-pressure stability region is completely contained inside the constant-volume stability region, indicating that constrained catenoids are more stable than unconstrained catenoids. }

\section{Concluding remarks} \label{sec:conclusion}
We have examined the equilibrium and stability of catenoids bridging a circular disc and plate where the equilibrium surface is pinned at one contact line to the disc edge with the other free to move on the plate. Drawing on the second variation of potential energy, the existence, maximal stability, and stability regions were analytically determined. The equilibrium solution multiplicity subject to various constraints was discussed in detail. The results showed that all catenoids are stable with respect to non-axisymmetric perturbations; for a fixed contact angle, there exists a critical volume below which catenoids are unstable to axisymmetric perturbations. The canonical phase diagram furnishes a one-to-one correspondence between points in the existence region and equilibrium solutions where the stability-region boundary separates the points corresponding to stable catenoids from those corresponding to unstable ones. No such correspondence can be established in the favourable phase diagram. Furthermore, the canonical phase diagram conveniently demonstrates how the catenoid geometrical symmetry affects the stability regions. For example, the maximal stability region symmetry with respect to the phase diagram minor diagonal indicates the reflective symmetry (with respect to the ratio of lower and upper disc diameters) of catenoids with two pinned contact lines. Moreover, the asymmetric shape of the stability region shows how a catenoid free contact line with a substrate breaks the equatorial and reflective symmetries. The stability limit presented here is a limiting case for the minimum volume stability limit of liquid bridges when the mean curvature approaches zero \citep{akbari2014bridge}. The static stability limits are useful for predicting the transition of the time scale from the quasi-static to the intermediate phases of contact-drop dispensing.

\section*{Acknowledgements} \label{sec:acknowledgements}

Supported by the NSERC Innovative Green Wood Fibre Products Network, and a McGill Engineering Doctoral Award (MEDA) to A.A.

\numberwithin{equation}{section}
\begin{appendices}
\section{Stability with respect to non-axisymmetric perturbations} \label{sec:appendixa}
We shall prove that all catenoids are stable with respect to non-axisymmetric perturbations. First, we show that $D^{1}(\hat{s}_{0},\hat{s}_{1})=0$ has no non-trivial root. Substituting  Eqs.~(\ref{eqn:eq44})-(\ref{eqn:eq48}) into Eq.~(\ref{eqn:eq26}) yields
\begin{equation}
	\hat{s}_{1}\sqrt{\hat{s}_{1}^2+1}+\ln(\hat{s}_{1}+\sqrt{\hat{s}_{1}^2+1})=\hat{s}_{0}\sqrt{\hat{s}_{0}^2+1}+\ln(\hat{s}_{0}+\sqrt{\hat{s}_{0}^2+1}),
    	 \label{eqn:eq50}
\end{equation}
where we seek an $\hat{s}_{0}\in(-\infty,\hat{s}_{1})$ for a fixed $\hat{s}_{1}$. Equation~(\ref{eqn:eq50}) is obviously satisfied for the trivial solution $\hat{s}_{0}=\hat{s}_{1}$. Denoting the left-hand side by $f(\hat{s}_{1})$, Eq.~(\ref{eqn:eq50}) can be rewritten as 
\begin{equation}
	g(\hat{s}_{0})=\hat{s}_{0}\sqrt{\hat{s}_{0}^2+1}+\ln(\hat{s}_{0}+\sqrt{\hat{s}_{0}^2+1})-f(\hat{s}_{1})=0
    	 \label{eqn:eq51}
\end{equation}
with
\begin{equation}
	\frac{dg}{d\hat{s}_{0}}=\sqrt{\hat{s}_{0}^2+1}+\frac{\hat{s}_{0}^{2}}{\sqrt{\hat{s}_{0}^2+1}}+\frac{1}{\sqrt{\hat{s}_{0}^2+1}}.
    	 \label{eqn:eq52}
\end{equation}
All the terms on the right-hand side of Eq.~(\ref{eqn:eq52}) are positive, indicating that $dg/d\hat{s}_{0}>0$ on $(-\infty,\hat{s}_{1})$. The continuity of $g(\hat{s}_{0})$ implies that $g$ is a monotonically increasing function such that $g\to-\infty$ as $\hat{s}_{0}\to-\infty$ and $g\to0^{-}$ as $\hat{s}_{0}\to\hat{s}_{1}$. Therefore, $\hat{s}_{0}=\hat{s}_{1}$ is the only solution of Eq.~(\ref{eqn:eq51}).

Next, we prove that $\hat{\chi}^{1}(\hat{s}_{0},\hat{s}_{1})-\hat{\chi}(\hat{s}_{1})=0$ has no non-trivial root. Substituting  Eqs.~(\ref{eqn:eq44})-(\ref{eqn:eq48}) into Eq.~(\ref{eqn:eq23}) results in
\begin{align}
	\frac{\hat{s}_{1}}{\hat{s}_{1}^2+1} \Biggl[ \frac{\frac{\sqrt{\hat{s}_{1}^2+1}(\hat{s}_{1}^2+2)}{\hat{s}_{1}}-\sinh^{-1}\hat{s}_{1}+\hat{s}_{0}\sqrt{\hat{s}_{0}^2+1}+\sinh^{-1}\hat{s}_{0}}{\hat{s}_{1}\sqrt{\hat{s}_{1}^2+1}+\sinh^{-1}\hat{s}_{1}-\hat{s}_{0}\sqrt{\hat{s}_{0}^2+1}-\sinh^{-1}\hat{s}_{0}}
	+1 \Biggr]=0.
	\label{eqn:eq53}
\end{align}
Note that the denominator of the fraction in the square bracket is non-zero for $\hat{s}_{0} \in (-\infty,\hat{s}_{1})$. Two casesmust be considered separately: (1) $\hat{s}_{1}\neq0$ and (2) $\hat{s}_{1}\to0$. The first leads to 
\begin{equation}
	 \sqrt{\hat{s}_{1}^2+1}+(\hat{s}_{1}^2+1)^{3/2}+\hat{s}_{1}^{2}\sqrt{\hat{s}_{1}^2+1}=0.
    	 \label{eqn:eq54}
\end{equation}
A pair $(\hat{s}_{0},\hat{s}_{1})$ that satisfies Eq.~(\ref{eqn:eq53}) must also satisfy Eq.~(\ref{eqn:eq54}). Equation~(\ref{eqn:eq54}) clearly shows that Eq.~(\ref{eqn:eq53}) has no non-trivial root since it is independent of $\hat{s}_{0}$. In addition, no $\hat{s}_{1}$ can be found that satisfies Eq.~(\ref{eqn:eq54}) because the left-hand side is always greater than zero. From the second case,
\begin{equation}
	 \frac{2}{\hat{s}_{0}\sqrt{\hat{s}_{0}^2+1}-\sinh^{-1}\hat{s}_{0}}=0,
    	 \label{eqn:eq55}
\end{equation}
which holds only when $\hat{s}_{0}\to -\infty$. This indicates that, for any given contact angle, only infinitely large catenoids may lose stability to non-axisymmetric perturbations, and the stability-region boundary coincides with the existence-region boundary in the canonical phase diagram. 

One can apply the same procedure as the two previous cases to prove that $D^{0}(\hat{s}_{0},\hat{s}_{1})=0$ has no non-trivial root for $\theta_{c}\leq \pi/2$. However, the expressions are cumbersome and the analysis is tedious. We only demonstrate the limiting behaviour discussed in section~\ref{sec:stability:catenoid}. The Taylor-series expansion of $D^{0}$ is used for the small-interface limit:
\begin{equation}
	 D^{0}=-\frac{\epsilon^{4}}{12}+O(\epsilon^{5}),\qquad \epsilon \ll1,
    	 \label{eqn:eq56}
\end{equation}
where $\epsilon=\hat{s}_{1}-\hat{s}_{0}$. It follows that
\begin{equation}
	 \lim_{\epsilon \to 0^{+}}D^{0}=0^{-}.
    	 \label{eqn:eq57}
\end{equation}
In the other limit, where catenoids are infinitely large, one can show that
 \begin{equation}
	 \lim_{\hat{s}_{0} \to -\infty}D^{0}=\mbox{sign}(\hat{s}_{1})\times \infty,
    	 \label{eqn:eq58}
\end{equation}
implying that there is at least one $\hat{s}_{0}\in(-\infty,\hat{s}_{1})$ at which $D^{0}(\hat{s}_{0},\hat{s}_{1})=0$ for $\theta_{c}>\pi/2$. These limits are clearly illustrated in Fig.~\ref{fig:figure5}.

\section{Symmetry of $D$-functions} \label{sec:appendixb}
We prove that $D^{0}$ and $D^{1}$ are symmetric with respect to the canonical phase diagram minor diagonal. Consider the following transformation
\begin{equation}
	\left\{ \begin{array}{l}
         		\hat{s}_{0}=-\bar{\hat{s}}_{1}\\
		\hat{s}_{1}=-\bar{\hat{s}}_{0}\\
	\end{array} \right. ,
    	 \label{eqn:eq59}
\end{equation}
which is equivalent to Eq.~(\ref{eqn:eq49}). Given that $w_{1}$, $w_{4}$ are odd and $r$, $w_{2}$, $w_{3}$, $w_{5}$ are even functions, we have
\begin{equation}
	D^{0}(\bar{\hat{s}}_{0},\bar{\hat{s}}_{1})=
    	\left| \begin{array} {c c c}
	-w_{1}(\hat{s}_{1}) & w_{2}(\hat{s}_{1}) & w_{3}(\hat{s}_{1})\\
	-w_{1}(\hat{s}_{0}) & w_{2}(\hat{s}_{0}) & w_{3}(\hat{s}_{0})\\
	-\int_{\hat{s}_{0}}^{\hat{s}_{1}} \hat{r} w_{1}d\hat{s} & \int_{\hat{s}_{0}}^{\hat{s}_{1}} \hat{r} w_{2}d\hat{s} & \int_{\hat{s}_{0}}^{\hat{s}_{1}} \hat{r} w_{3}d\hat{s}
	\end{array} \right|,
	 \label{eqn:eq60}
\end{equation}
\begin{equation}
	D^{1}(\bar{\hat{s}}_{0},\bar{\hat{s}}_{1})=
    	\left| \begin{array} {c c }
	-w_{4}(\hat{s}_{1}) & w_{5}(\hat{s}_{1})\\
	-w_{4}(\hat{s}_{0}) & w_{5}(\hat{s}_{0})
	\end{array} \right|.
	 \label{eqn:eq61}
\end{equation}
Taking the determinant row exchange rules into consideration, it follows that
\begin{equation}
	D^{0}(\bar{\hat{s}}_{0},\bar{\hat{s}}_{1})=D^{0}(\hat{s}_{0},\hat{s}_{1}),
  	 \label{eqn:eq62}
\end{equation}
 \begin{equation}
	D^{1}(\bar{\hat{s}}_{0},\bar{\hat{s}}_{1})=D^{1}(\hat{s}_{0},\hat{s}_{1}).
  	 \label{eqn:eq63}
\end{equation}
This completes the proof.

\end{appendices}

\bibliography{mybib}
\bibliographystyle{unsrtnat}

\end{document}